\documentclass[amsmath,amssymb,aps,prl,preprint]{revtex4-2}

\usepackage{graphicx}
\usepackage{subfig}

\newcommand{\mr}[1]{\mathrm{#1}}

\begin{document}

\title{Molecular Structure Optimization based on Electrons-Nuclei Quantum Dynamics Computation}

\author{Hirotoshi Hirai}
\email{hirotoshih@mosk.tytlabs.co.jp}
\author{Takahiro Horiba}
\author{Soichi Shirai}

\affiliation{Toyota Central R\&D Labs., Inc., 41-1, Yokomichi, Nagakute, Aichi 480-1192, Japan}

\author{Keita Kanno}
\author{Keita Arimitsu}
\author{Yuya O. Nakagawa}
\author{Sho Koh}

\affiliation{QunaSys Inc., Aqua Hakusan Building 9F, 1-13-7 Hakusan, Bunkyo, Tokyo 113-0001, Japan}

\date{\today}

\begin{abstract}
A new concept of the molecular structure optimization method based on quantum dynamics computations is presented.
Nuclei are treated as quantum mechanical particles, as are electrons, and the many-body wave function of the system is optimized by the imaginary time evolution method. 
A demonstration with a 2-dimensional H$_2^+$ molecule shows that the optimized nuclear positions can be specified with a small number of observations.
This method is considered to be suitable for quantum computers, the development of which will realize its application as a powerful method. 
\end{abstract}

\maketitle

%\section{Introduction}
{\it Introduction.}
Accurate quantum chemistry computations remain a challenge on classical computers, especially for molecules with industrially relevant sizes, despite significant efforts and developments of recent years.
The computational cost of exact methods for quantum chemistry on classical computers grows exponentially with the molecular size~\cite{KNOWLES1984, JOlsen1988}, whereas that can be suppressed in polynomial scaling on quantum computers~\cite{Nielsen2011}.
For this reason, quantum chemistry computations have been considered to be a promising application of quantum computers.
By manipulating quantum states of matter and taking advantage of their unique features, such as superposition and entanglement, quantum computers promise to efficiently deliver accurate results for many important problems in quantum chemistry, such as the electronic structure of molecules \cite{zalka1998simulating, zalka1998efficient}.

In quantum chemistry computations, the Born-Oppenheimer (BO) approximation \cite{combes1981born} has typically been used to speed up the computation of molecular wave functions and other properties for large molecules.
The BO approximation is the assumption that the wave functions of atomic nuclei and electrons in molecular systems can be treated separately, based on the fact that the nuclei are much heavier than the electrons.
In many cases of quantum chemistry computations by classical computers, nuclei have been treated as point charges (classical particles).
Many quantum algorithms for quantum chemistry computations on quantum computers are also typically based on the BO approximation: for example, the quantum phase estimation (QPE) method~\cite{kitaev1995quantum, Cleve1998} for a fault-tolerant quantum computer (FTQC) %\cite{wang201816}
and the variational quantum eigensolver (VQE) method \cite{Peruzzo2014,kandala2017hardware} for a noisy intermediate-scale quantum (NISQ) \cite{preskill2018quantum} device are often employed to solve systems under the BO approximation.
On the other hand, it is also possible to think of methods beyond the BO approximation with quantum computers, and such methods have been proposed in literature~\cite{zalka1998simulating, Stieve1998, kassal2008polynomial, ollitrault_non-adiabatic_2020}.
Kassal et al. \cite{kassal2008polynomial}, for instance, reported that a completely non-adiabatic grid-based method on FTQCs, where nuclei are treated as quantum particles and correlated with electrons, is not only more accurate but also faster and more efficient than the methods based on the BO approximation.

Geometry optimization of molecules is an important process to obtain the  equilibrium molecular structures in quantum chemistry computations \cite{eckert1997ab}.
Since the physical and chemical properties of molecules are dependent on their specific geometrical structures, elucidation of the optimized structures enables the prediction of properties and identification of chemical products.
In the conventional molecular structure optimization methods, which are based on the BO approximation and treat nuclei as classical particles, the electronic states should be computed to obtain the forces acting on nuclei and update the molecular structure for each optimization step.
These conventional methods can be accelerated through the use of quantum computers to perform the electronic state calculations.

However, a large number of measurements (circuit executions) are required because it is necessary to repeat the electronic state computations on the quantum computer at each iteration in the geometry optimization process.
Furthermore, as gradient-driven descent routines are often used to optimize the molecular structure, the system tends to be relaxed to the closest local minimum from an initial structure for the geometry optimization.
It is thus a difficult task to find the global minimum structure in the systems which have multiple local minima (i.e. many isomers), such as alloy cluster systems \cite{eckert1997ab}.

In this study, we propose a method for optimizing molecular structures based on quantum dynamics computations with working on an FTQC in mind.
In our method, the many-body wave functions of nuclei and electrons are directly treated as wavepackets and optimized by the imaginary time evolution method.
The many-body wave function on the lattice can be mapped to a state on a quantum computer with the number of qubits growing only logarithmically with respect to the number of lattice points \cite{Stieve1998,zalka1998efficient}.
Although the imaginary time evolution is a non-unitary operation, probabilistic methods for FTQC are proposed \cite{williams2004probabilistic,terashima2005nonunitary,liu2021probabilistic}, and the method of linear combination of unitaries could be also be useful
\cite{childs2012hamiltonian,berry2015simulating}. A variational method for NISQ devices is also proposed \cite{mcardle2019variational}. We leave a concrete implementation of our proposal on quantum computers as a future work.
Since nuclei are treated as quantum particles, the nuclear quantum effects are naturally included in this method.
If we start with an initial many-body wave function that covers all molecular structures, it is expected theoretically that the global minimum of the molecular structure can be obtained. 
The optimized many-body wave functions have large stochastic amplitudes at the most stable structure; therefore, the nuclei positions can be specified, i.e. the optimization of the molecular structure is possible, with a small number of observations.
We demonstrate the concept of the molecular structure optimization method based on quantum dynamics computations with a classical computer and the size of the molecules that can be handled.

{\it Quantum imaginary time evolution.}
In non-relativistic quantum dynamics computations, the time evolution of quantum systems can be described by the time-dependent Schr\"{o}dinger equation:
\begin{equation}
i\hbar\frac{\partial}{\partial t}\psi(t) = H \psi (t),
\end{equation}
where $H$ is the Hamiltonian that describes the system.
The solution of the time-dependent  Schr\"{o}dinger equation is formally written by
\begin{equation}
\psi(t) = e^{-i \frac{H}{\hbar} t}\psi (0).
\end{equation}
That is, if an appropriate initial wave function $\psi (0) $ can be prepared, then the time evolution of the system can be obtained by application of the time evolution operator,
\begin{equation}
U(t) = e^{-i \frac{H}{\hbar} t}.
\end{equation}
If $\psi (0)$ is the eigenstate of $H$, i.e. $\psi(0)=\psi_i$ and $H\psi_i=E_i\psi_i$, then Eq. (3) becomes 
\begin{equation}
U(t)\psi(0) = e^{-i \frac{E_i}{\hbar} t}\psi_i.
\end{equation}
If not (in general cases), then the exponential operator must be applied to the wave function to compute the time evolution of the system.
A number of methods have been developed to apply the exponential operator to the wave function~\cite{leforestier1991comparison, van2007accurate}. We here explain the method based on the second-order Suzuki-Trotter decomposition~\cite{zalka1998efficient, zalka1998simulating, barthel2020optimized}.

Denoting the kinetic energy term of $H$ by $T$ and the potential energy term by $V$, the second-order Suzuki-Trotter decomposition is 
\begin{equation}
U(t) =e^{-i \frac{T}{\hbar} \frac{t}{2}}e^{-i \frac{V}{\hbar} t}e^{-i \frac{T}{\hbar} \frac{t}{2}}
+O(t^3).
\end{equation}
The error arises because $T$ and $V$ are non-commutative. 
To reduce the error, we set $t = N_{\mr{step}} dt$ and express the time evolution operator as the product of $N_{\mr{step}}$ operators in time increments of $dt$,
\begin{equation}
U(t) = (U(dt))^{N_\mr{step}} = (e^{-i \frac{H}{\hbar} dt})^{N_\mr{step}}.
\end{equation}
Because $T$ is diagonal in wavenumber space,
when we apply $e^{-i \frac{T}{\hbar} \frac{dt}{2}}$ in the second-order Suzuki-Trotter decomposition, the wave function is expressed in wavenumber space by performing the fast Fourier transformation (FFT) for the wave funciton in real space, when running on classical computers.
Similarly, $V$ is diagonal in real space, so the operator $e^{-i \frac{V}{\hbar} dt}$ is applied to the wave function in real space after application of the inverse FFT to the wave function in wavenumber space, $e^{-i \frac{T}{\hbar}\frac{dt}{2}}\psi$. 
Using these procedures, the application of the time evolution operator $U(dt)$ can be computed without expanding the exponential function operator.
By repeating these calculations $N_\mr{step}$ times, the wave function at the desired time $t$ can be obtained. 

%\subsection{Imaginary time evolution}
Now let us convert time into an imaginary number, $it\rightarrow \tau$.
The solution of the time-dependent Schr\"{o}dinger equation then becomes
\begin{equation}
\psi(\tau) = e^{-\frac{H}{\hbar} \tau}\psi (0).
\end{equation}
When we formally expand $\psi(0)$ by the eigenfunctions of $H$, $\{\phi_i\}_i$, we have
\begin{equation}
\psi(\tau) = \sum_i c_i e^{-\frac{E_i}{\hbar} \tau}\phi_i,
\end{equation}
where $c_i \in \mathbb{C}$ is a coefficient for the expandsion $\psi(0)=\sum_i c_i \phi_i$, and $E_i$ is the eigenvalue of $\phi_i$.
If we increase $\tau$, the factor $e^{-\frac{E_i}{\hbar} \tau}$ attenuates faster for large $E_i$, and the ground state $\phi_0$, which has the smallest eigenvalue $E_0$, remains until the end as long as the coefficient (overlap between the initial state $\psi(0)$ and the ground state) $c_0$ is non-zero.
This is how we obtain the ground state by imaginary time evolution.
Note that the norm of the wave function changes during the imaginary time evolution, so it is necessary to renormalize the wave function after the evolution.

{\it Geometry optimization with fully quantum wave function for nuclei and electrons.}
In this study, in contrast to the conventional method based on the BO approximation, we propose a geometry optimization method where the atomic nuclei are regarded as quantum mechanical particles as well as the electrons and treated by wave functions.
We consider the many-body wave function $\psi(R, r, \tau=0)$, including the degrees of freedom of nuclei and electrons, where $R$ represents the coordinates of $N_\mr{nuc}$ nuclei,
$R={R_1, R_2, ... , R_{N_\mr{nuc}}}$,
and $r$ represents the coordinates of $n$ electrons,
$r={r_1, r_2, ... , r_n}$.
Starting from any initial wave function (ultimately, it may be a constant over the entire simulation region in real space, but should have non-zero overlap with the desired ground state), 
the most stable state (ground state) of the nuclei and the electrons can be obtained by the imaginary time evolution of the initial wave function under the Hamiltonian 
\begin{equation}
\begin{split}
H &= -\sum_i^n \frac{\hbar^2}{2m_i}\nabla_{r_i}^2
-\sum_i^{N_\mr{nuc}} \frac{\hbar^2}{2M_i}\nabla_{R_i}^2 + \sum_{i<j}\frac{q_e^2}{|r_i-r_j|}\\
&+ \sum_{i<j}\frac{Q_i Q_j}{|R_i-R_j|} - \sum_i^{N_\mr{nuc}} \sum_j^n \frac{q_e Q_i}{|R_i-r_j|},
\end{split}
\end{equation}
where $Q_i$ is a electric charge of $i$th nuclear and $q_e (>0)$ is a charge of electron.
The optimized wave function for the nuclei, which can be obtained by integrating out the electronic degrees of freedom, does not spread like electrons and has sharp peaks (cusps) at the most stable positions; therefore, the position of each nuclei should be identifiable with several repetitions of the projective measurement for the positions of the nuclei.
The stable structure can be obtained with a high probability by performing the measurement to the state after imaginary time evolution. 
This calculation includes quantum effects such as the zero-point oscillation effect and the non-adiabatic effects that are ignored when the atomic nuclei are treated as point charges \cite{nakai2007nuclear}.
If we take an initial state (for the nuclei part of the wave function) as the superposition state for all possible molecular structures, the most stable structure among all isomers of the molecule would be obtained. 
Because the search for the most stable isomer generally becomes difficult as the number of atoms increases \cite{eckert1997ab}, it is an appealing feature of this method that treats the atomic nuclei as quantum mechanical particles.

Such an approach could not be taken to date due to the difficulty of describing many-body wave functions with a classical computer.
In contrast, a many-body wave function can be described with a decent number of qubits on a quantum computer. This is discussed in more detail later.

{\it Numerical demonstration of proof-of-concept for our method.}
We present numerical demonstrations of our method by taking a two-dimensional (2D) H$_2^+$ molecule as an example. 
This is a three-particle system with two protons and an electron that is naively described as a six-dimensional system, which will be difficult to calculate with a classical computer. 
For example, if we take a grid of $2^5 = 32$ per dimension in that space, the wave function has $2^{30} = 1,073,741,824$ elements, and the Hamiltonian has $2^{60} = 1,152,921,504,606,846,976$ elements in a naive estimate. 
On the other hand, in a quantum computer, a wave function can be expressed with 30 qubits because $N$ qubits can represent $2^N$ variables, i.e. $2^N$ wave function elements on grids. In addition, the number of qubits scales linearly with the number of particles. 

In numerical demonstrations, we perform a dimensional reduction to the system so that the calculation can be performed on a classical computer. 
First, the center of mass of the system is fixed at the origin $(0, 0)$. 
As a result, the degrees of freedom of translational motion of the molecule can be ignored. 
Now that the purpose is to optimize the molecular structure, we only need to focus on the internal degrees of freedom. 
Classically, the rotational motion around the center of mass has also been frozen, but quantum mechanically, the rotational motion and the vibration mode are coupled, so that the rotational motion cannot be separated. 
Even if the rotational motion is left, observing the optimized wave function gives only one molecular structure in a specific orientation of the rotational degrees of freedom. 
When we obtain multiple results for the measurements of the positions of the nuclei, the obtained structures are described in terms of internal coordinates which allows rotational-free description of molecules. 
The mean of each nucleus position is then calculated, which gives the molecular structure.
Next, the relative coordinates $\vec{R}$ between the two nuclei are introduced to further reduce the dimension. 
The relationship between the reduced coordinates and the original coordinates is given as 
\begin{equation}
\vec{R} = \vec{r}_2 - \vec{r}_1,
\: \vec{r}_c = \vec{r}_e - \frac{m_1\vec{r_1}+m_2\vec{r_2}}{M_n},
\end{equation}
\begin{equation}
\vec{R}_\mr{cm} = \frac{m_1\vec{r_1}+m_2\vec{r_2}+m_e\vec{r_e}}{M} \: (\rightarrow \text{set to } (0, 0)),
\end{equation}
where $\vec{r}_1$ and $\vec{r}_2$ ($m_1$ and $m_2$) are the coordinates (masses) of the two protons, 
$\vec{r}_e$ ($m_e$) are the coordinates (mass) of the electron, 
$\vec{r}_c$ are the coordinates of the electron as seen from the center of mass of the nuclei, and $\vec{R}_\mr{cm}$ is the center of mass of the system and fixed at the origin.
We here also define the total mass of the system $M = m_1 + m_2 + m_e$ and that of the protons $M_n = m_1 + m_2$.
The reduced coordinates have four degrees of freedom, $\vec{R}=(R_x, R_y)$ and  $\vec{r}_c=(r_{cx}, r_{cy})$.
The reduced masses for $\vec{R}$ and $\vec{r}_c$ are
$ \mu_R^{-1} = m_1^{-1} + m_2^{-1}$
and
$\mu_{r_c}^{-1} = m_e^{-1} + M_n^{-1}$, respectively.
By using the reduced coordinates, the Hamiltonian of the system becomes
\begin{equation}
\begin{split}
H &= - \frac{1}{2\mu_R}\nabla_R^2  - \frac{1}{2\mu_{r_c}}\nabla_{r_c}^2\\
&+ \frac{1}{|\vec{R}|} - \frac{1}{|\vec{r}_c+\frac{m2}{M_n}\vec{R}|} - \frac{1}{|\vec{r}_c-\frac{m1}{M_n}\vec{R}|}.
\end{split}
\end{equation}
Here the atomic unit is introduced for simplicity ($\hbar = 1$, $Q_p = 1$, $q_e = 1$).
We take an evenly spaced isotropic grid of $2^5$ elements per dimension in each two-dimensional spaces for $\vec{R}, \vec{r}_c$.
See Supplemental Material for more details on computational setup.

\begin{figure}[htp]
\centering
\includegraphics[width=9.5cm]{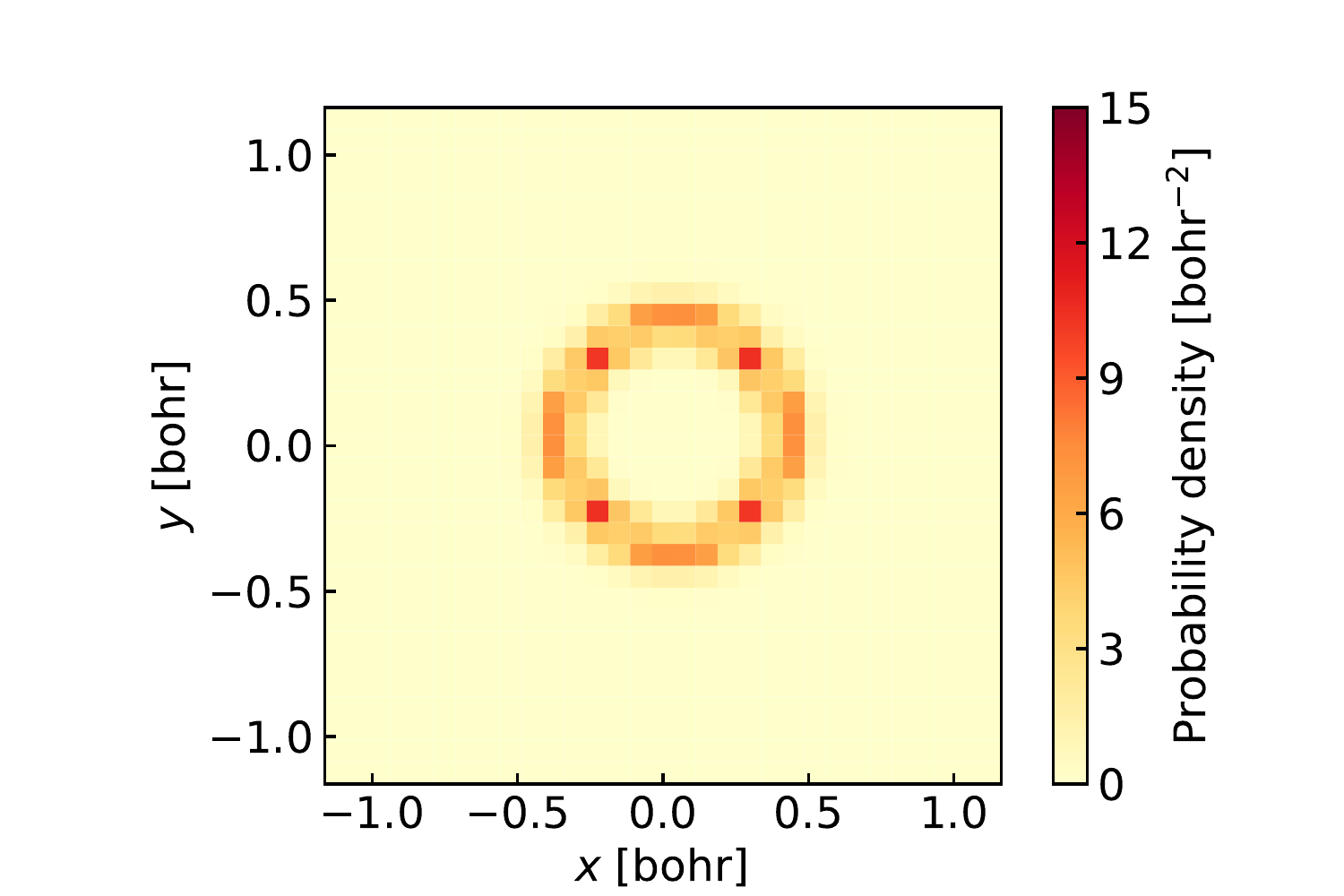}
\caption{Conditional probability density distribution of the relative coordinates of protons $\vec{R}$ for the H$_2^+$ molecule when an electron is near the origin point (0.0375, 0.0375).}
\label{fig1}
\end{figure}

%\section{Results \& Discussions}
We present results of numerical demonstrations result for the method using the 2D H$_2^+$ molecule. 
The conditional probability density distribution of the relative distance $\vec{R}$ between the protons when the electron is near the origin (0.0375, 0.0375) (note that the grid used in this study does not contain (0, 0)), $|\psi(R_x, R_y, r_{cx}=0.0375, r_{cy}=0.0375)|^2$ is shown in FIG. \ref{fig1}. 
We find that the probability density distribution of the relative coordinates $\vec{R}$ between the protons has a donut-shaped structure. 
This means that the H$_2^+$ molecule can be oriented in any direction because it has rotational degrees of freedom. 
This probability density distribution must be donut-shaped and uniform for all directions; however, the probability is large and small in specific directions in FIG.~\ref{fig1}. 
The reason for this is considered to be that the roughness of the grid space used in this study cannot represent an appropriate distance for the equilibrium internuclear distance in specific directions and the probability density distributions for such directions would be small, and the probability density distributions for the specific directions where the grid is well-matched with the equilibrium internuclear distance would be large.
We also calculate the expectation value (mean) of the proton-proton distance $|\vec{R}|$, defined as
\begin{equation}
\int |\vec{R}| |\psi(\vec{R},\vec{r}_c)|^2 d\vec{R}d\vec{r}_c,
\end{equation}
for our simulation data in FIG.~\ref{fig1}.
The result is 0.41 bohr, which is slightly larger than the value of 0.37 bohr where the protons were treated as point charges (see Supplemental Material). 
This larger expectation value is considered to be due to the spread of the nuclear wave function and its binding potential, which is very tight for smaller $|\vec{R}|$ but loose for larger $|\vec{R}|$.

Next, we estimate how many measurements for the wave function would be required to optimize the molecular structure within an acceptable error. 
$|\psi(\vec{R}, \vec{r_c})|^2$ gives the simultaneous probability of finding relative proton-to-proton coordinates in $\vec{R}$ and electrons in $ \vec{r_c} $. 
To optimize the molecular structure, it is enough to measure the positions of the nuclei coordinates ($\vec{R}$ in this case).
We simulate the measurements (observations) of $\vec{R}$ for the optimized wave function by randomly sampling $\vec{R}$ with the probability $p(\vec{R}) = \int |\psi(\vec{R}, \vec{r_c})|^2 d\vec{r}_c$ using pseudo-random numbers generated by the Mersenne Twister method~\cite{MT1998}.
\begin{figure}[htp]
\centering
\includegraphics[width=7.5cm]{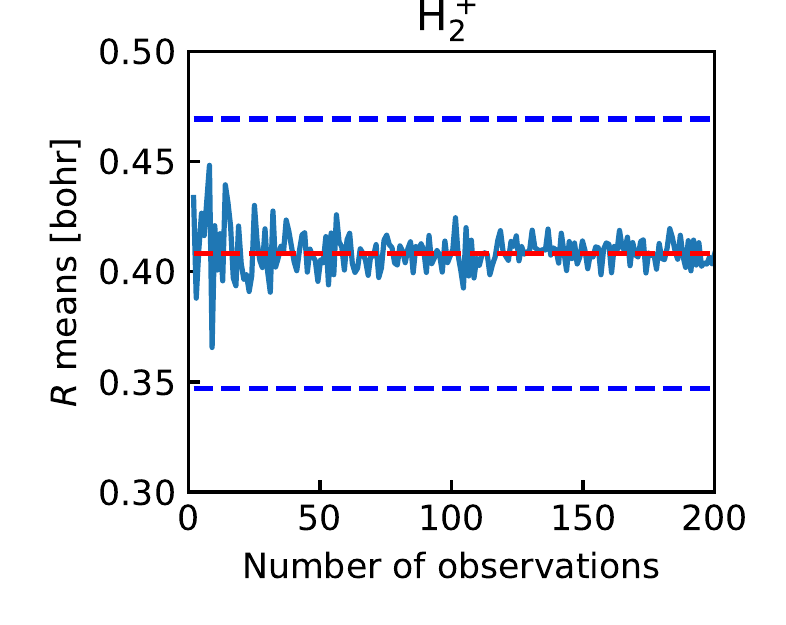}
\caption{
Mean of the sampled bond distance $|\vec{R}|$ drawn from the optimized wave function.
We plot the number of observations $N_\mr{obs}$ versus the mean up to $N_\mr{obs}$ observations, $M(N_\mr{obs}) = \frac{1}{N_\mr{obs}}\sum_{i=1}^{N_\mr{obs}} |\vec{R}_i|$, where $\vec{R}_i$ is an $i$-th sample of the observed $\vec{R}$.
The red (blue) dotted line represents the exact mean (standard deviation) of $|\vec{R}|$.
}
\label{fig2}
\end{figure}
Figure~\ref{fig2} shows how the mean of $|\vec{R}|$ for $N_\mr{obs}$ observations converges as we increase $N_\mr{obs}$.
The exact mean and the standard deviation (corresponding for the zero-point vibration of the ground state) calculated from $p(\vec{R})$ are also shown in dotted lines.
We see that the mean of the sampled $|\vec{R}|$ gets close to the exact mean with sufficient accuracy after as small as 200 observations.
The mean $|\vec{R}|$ obtained from 200 observations was 0.4082 bohr while exact one is 0.41 bohr as explained earlier.
The standard deviation for 200 observations was 0.0611 bohr and the mode was 0.4142 bohr.
The reason why the converged bond length $\vec{R}$ was obtained with such a small number of observations is that the wave function has a peak (cusp) at the most stable nuclei positions and that the measurements will pick it up with high probability. 
It should be noted that as the size of the molecule increases, the number of vibrational modes also increases.
As a result, the fluctuation width of the zero-point vibration also increases, which may affect the required number of measurements to determine the molecular structure with a certain precision.
As discussed in the Supplemental Material, the standard deviation of the zero-point fluctuation scales as $O(\sqrt{N})$, where $N$ is the number of nuclei.
Thereby the required number of measurements for a certain precision will scale as $O(N)$.

\begin{figure}[htp]
\centering
\subfloat[D$_2^+$ molecule]{\includegraphics[width=9.5cm]{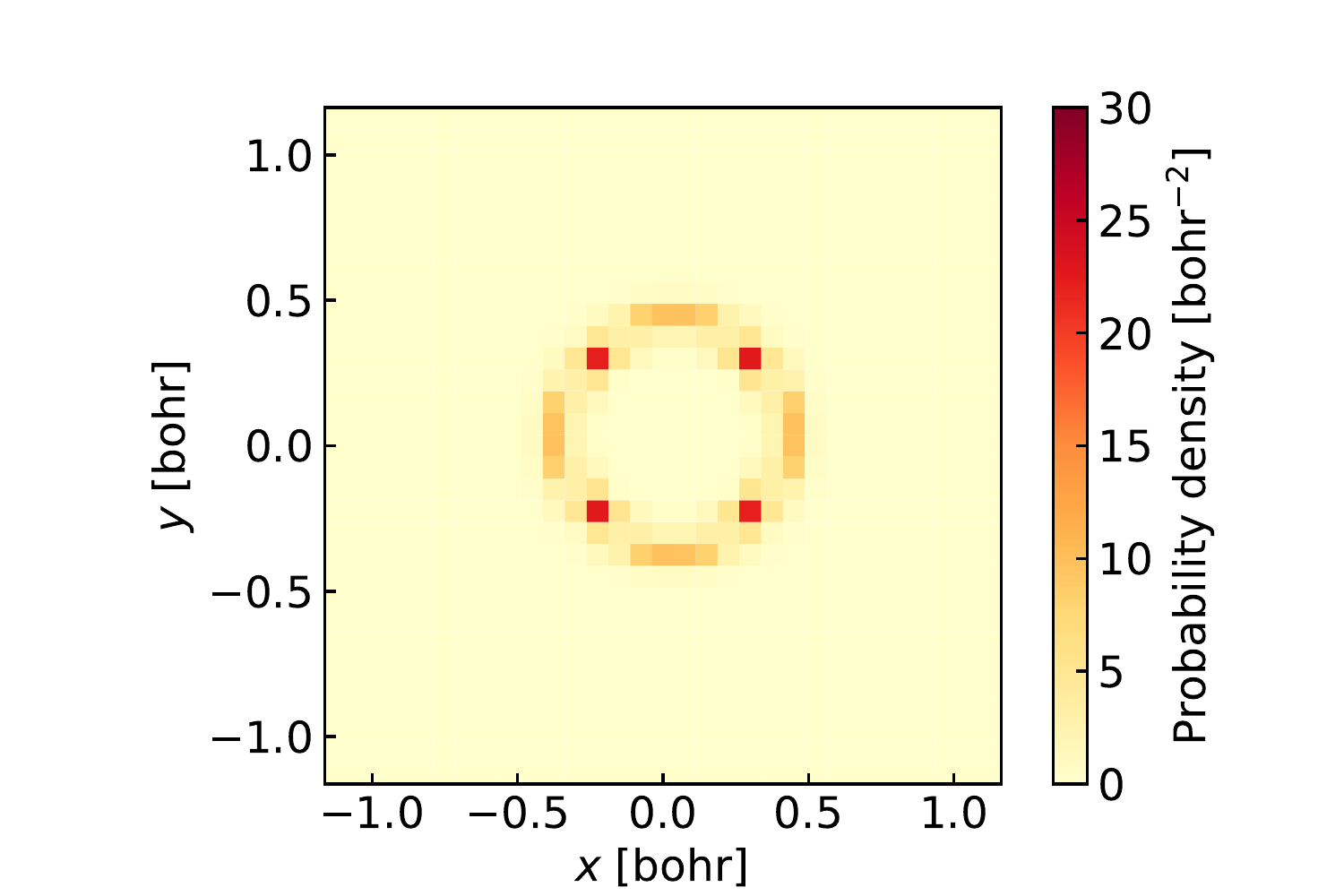}}
\\
\subfloat[T$_2^+$ molecule]{\includegraphics[width=9.5cm]{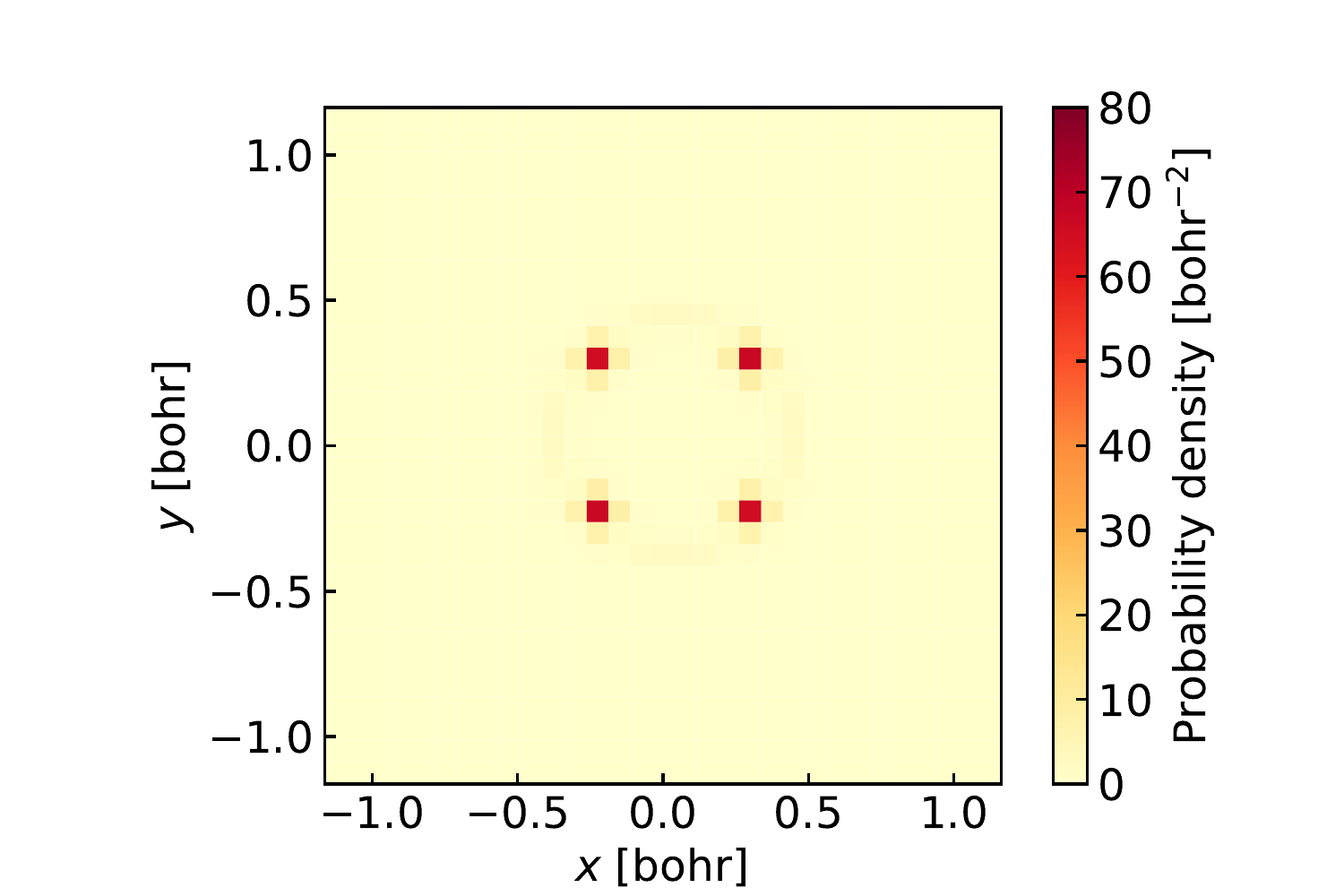}}
\caption{Conditional probability density distributions of the relative coordinates of nuclei for (a) D$_2^+$ and (b) T$_2^+$ molecules when the electron is near the origin point (0.0375, 0.0375).}
\label{fig3}
\end{figure}

\begin{figure}[htp]
\centering
\subfloat[D$_2^+$ molecule]{\includegraphics[width=7.5cm]{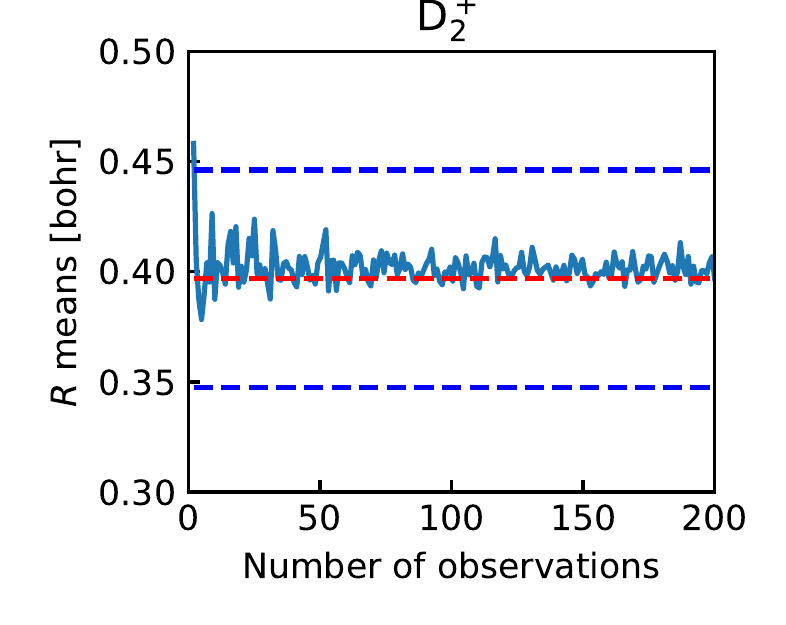}}
\\
\subfloat[T$_2^+$ molecule]{\includegraphics[width=7.5cm]{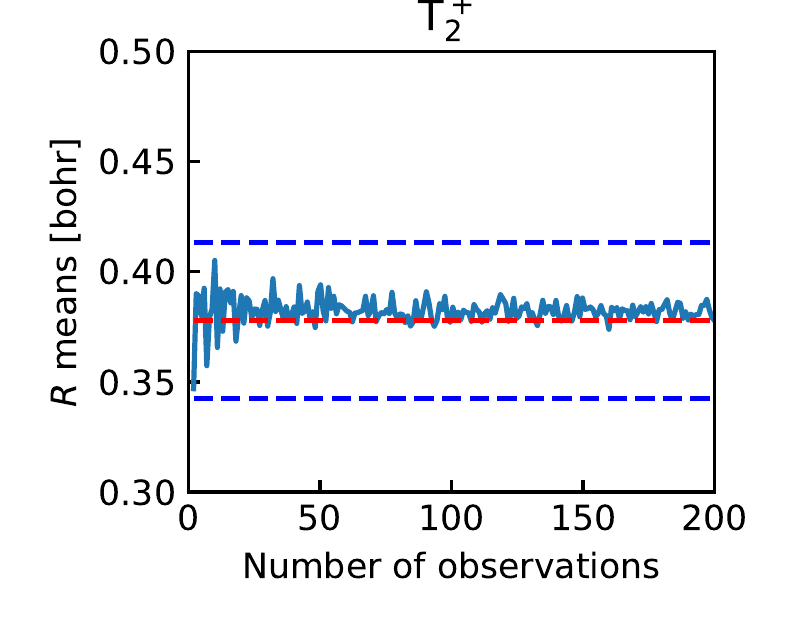}}
\caption{
Mean of the sampled bond distance $|\vec{R}|$ drawn from the optimized wave function for (a) D$_2^+$ and (b) T$_2^+$ molecules.
We plot the number of observations $N_\mr{obs}$ versus the mean up to $N_\mr{obs}$ observations, $M(N_\mr{obs}) = \frac{1}{N_\mr{obs}}\sum_{i=1}^{N_\mr{obs}} |\vec{R}_i|$, where $\vec{R}_i$ is an $i$-th sample of the observed $\vec{R}$.
The red (blue) dotted line represents the exact mean (standard deviation) of $|\vec{R}|$.
}
\label{fig4}
\end{figure}

Finally, we illustrate how the presented result above changes when we consider heavier nuclei.
It is expected that as a nucleus becomes heavier, the less wave-like properties it has and the closer it is to a classical particle.
It is predicted that the peaks at the nuclear positions of the wave function will be even sharper and converged structures of molecules will be obtained with a smaller number of observations. 
To see this, we consider the deuterium molecular ion D$_2^+$, in which the protons in the hydrogen molecular ion are replaced with deuterons (the mass is twice that of proton), and 
the tritium molecular ion T$_2^+$, in which tritons (the mass is three times that of proton) are substituted.
We perform the same calculations as for H$_2^+$ for these isotopes. 
Figure~\ref{fig3} show the conditional probability density distributions of the relative coordinates $\vec{R} $ for deuterons and tritons, respectively, when the electron is near the origin (0.0375, 0.0375). 
The absolute values of the peaks are higher than those for H$_2^+$,
and the spreads of the wave functions are narrowed. 
This tendency was more significant for the heavier T$_2^+$.
In addition, the peak bias increases for specific directions with the heavier isotopes D$_2^+ $ and T$ _2^+$ because as the wave functions become narrower, it becomes difficult for the grid points to exist at the positions where the peaks in the continuous limit (infinite number of meshes of the gird) because the grid size is not fine enough. 
Subsequently, as in the case of H$_2^+$, we simulate how many observations were required to optimize the molecular structure of D$_2^+ $ and T$ _2^+$ within an acceptable error. 
Figure~\ref{fig4} plots the same quantity as Fig.~\ref{fig2}, the mean of the sampled $|\vec{R}|$ from the optimized wave function, for D$_2^+ $ and T$_2^+$ molecules.
The exact mean and the standard deviation are also shown in dotted lines.
Besides, the means, standard deviations, and modes obtained from 200 observations are summarized in TABLE. \ref{table1} along with the result for H$_2^+$ for comparison. 
\begin{table}
  \caption{Mean, standard deviation and mode of the 200 times observation for $|\vec{R}|$ in bohr.}
  \label{table1}
  \centering
  \begin{tabular}{c|c|c|c}
    \hline
    System  & Mean  &  Standard Deviation  & Mode\\
    \hline
    H$_2^+$  & 0.4082  & 0.0611 & 0.4142 \\
    D$_2^+$  & 0.3968  & 0.0494 & 0.3712 \\
    T$_2^+$  & 0.3778  & 0.0355 & 0.3712 \\
    \hline
  \end{tabular}
\end{table}
These results in FIG. \ref{fig4} and TABLE. \ref{table1} inidicate that the variation of the observed $\vec{R}$ becomes smaller and the convergence becomes faster as the mass increases.
This is because of the smaller fluctuation widths of the zero-point vibrations for heavier nuclei.
Since all the nuclei after the He atom in the periodic table are heavier than triton, it is expected that molecular structures obtained by measurements will converge even faster in term of the number of the measurements. 
We briefly comment on the bond length $|\vec{R}|$ before ending the explanation of numerical demonstrations.
The mean of the bond length approaches the value for point charges (0.37 bohr) as the mass increases. 
The tendency for the bond length to become shorter as the mass increases has also been confirmed for the 3D hydrogen molecule and its isotope molecules \cite{mielke2002hierarchical}. 
The modes of the sampled $|\vec{R}|$ are the same for D$_2^+$ and T$_2^+$ because the values between 0.3712 and 0.4142 cannot be taken due to rough grid used in our calculations.
This result indicates that the step size of the grid must be sufficiently smaller than the desired resolution of the molecular structures. 
For calculations on a classical computer, making the grid finer leads to a large increase in computational cost, but on a quantum computer it will scales in polynomial time~\cite{kassal2008polynomial}.

{\it Summary and outlook.}
In summary, the concept of a molecular structure optimization method using quantum dynamics computation has been presented. 
The nuclei have been treated as quantum mechanical particles as were the electrons, and the many-body wave function of the system was optimized using the imaginary time evolution method. 
The optimized wave function has a large probability amplitude at the most stable structure of the nuclei, which allows us to determine the optimized nuclear positions with a small number of observations. 
It is expected that our method has a favorable feature that the most stable isomer structure can be obtained for complex systems without being trapped in local minima by virtue of the imaginary time evolution.
This method may be particularly promising for searching for the most stable structure for systems with many isomers, such as metal alloy clusters. 
Ultimately, it may also be useful to solve protein folding problems. 
Another aspect of our method is that it includes nuclear quantum effects (zero-point oscillation, non-adiabatic correction, etc.) that are typically ignored by conventional methods. 
The coulomb interactions between multiple particles are explicitly incorporated and there is no approximation in nucleus-nucleus, electron-nucleus and electron-electron interactions. 
Although executing our method for industrially-interesting large molecules with classical computers is difficult because of huge computational costs in treating the fully quantum wave function for nuclei and electrons,
quantum computers (possibly FTQC) can be an appealing candidate to run our method.
Our proposal can give a new insight for quantum chemistry computations on quantum computers.

\bibliographystyle{unsrt}
\bibliography{ref}
\end{document}

\title{Supplemental Material for ``Molecular Structure Optimization based on Electrons-Nuclei Quantum Dynamics Computation"}

\author{Hirotoshi Hirai}
\email{hirotoshih@mosk.tytlabs.co.jp}
\author{Takahiro Horiba}
\author{Soichi Shirai}

\affiliation{Toyota Central R\&D Labs., Inc., 41-1, Yokomichi, Nagakute, Aichi 480-1192, Japan}

\author{Keita Kanno}
\author{Keita Arimitsu}
\author{Yuya O. Nakagawa}
\author{Sho Koh}

\affiliation{QunaSys Inc., Aqua Hakusan Building 9F, 1-13-7 Hakusan, Bunkyo, Tokyo 113-0001, Japan}

\date{\today}

\maketitle

\section{Computational details for quantum dynamics computations}
In numerical demonstrations in the main text, the 2D space was represented by an evenly spaced isotropic grid of $2^5\times2^5 = 32\times32$. 
The step sizes of $dx$ and $dy$ were set to 0.075 bohr, and the ranges of the x-axis and y-axis for $\vec{R}$ and $\vec{r}_c$ were (-1.1625: 1.1625) bohr. 
The wave function of the system is given by two variables of $\psi (\vec{R}, \vec{r_c})$; therefore, the number of elements of the wave function is $(2^5)^{2} \times (2^5)^{2} = 2^{20}$. 
There are singular points where the Coulomb interaction term diverges to $\pm \infty$, and we replace the value $\pm \infty$ by a finite value $\pm \frac{1}{\sqrt{dx^2 + dy^2}}$. 
The initial state of the wave function is set to a Gaussian function centered on the origin,
\begin{equation}
\psi(\vec{R}, \vec{r_c}, \tau=0) = \frac{1}{(\alpha\pi)^2}e^{-\frac{R^2_x+R^2_y+r_{cx}^2+r_{cy}^2}{2\alpha^2}},
\end{equation}
where $\alpha = 0.5$ was used.
The second-order Suzuki-Trotter decomposition method was used to calculate the quantum dynamics of the system in the imaginary time, and the time step $\tau$ of the imaginary time evolution was set to 0.075 a.u. (1.81416 attoseconds). 
The mass of the proton is set to 1863.15, and when calculating the deuterium ion molecule D$_2^+$ and the triton ion molecule T$_2^+$ to confirm the isotope effect, twice or three times that mass was used. 
The mass of an electron is 1 in the atomic unit. 
In this study, it is assumed that the overlap of wave functions between nuclei is almost zero, and there is no exchange interaction between them. 

\section{Calculation for point charge particles (classical particles)}
\begin{figure}
\centering
\includegraphics[width=8.6cm]{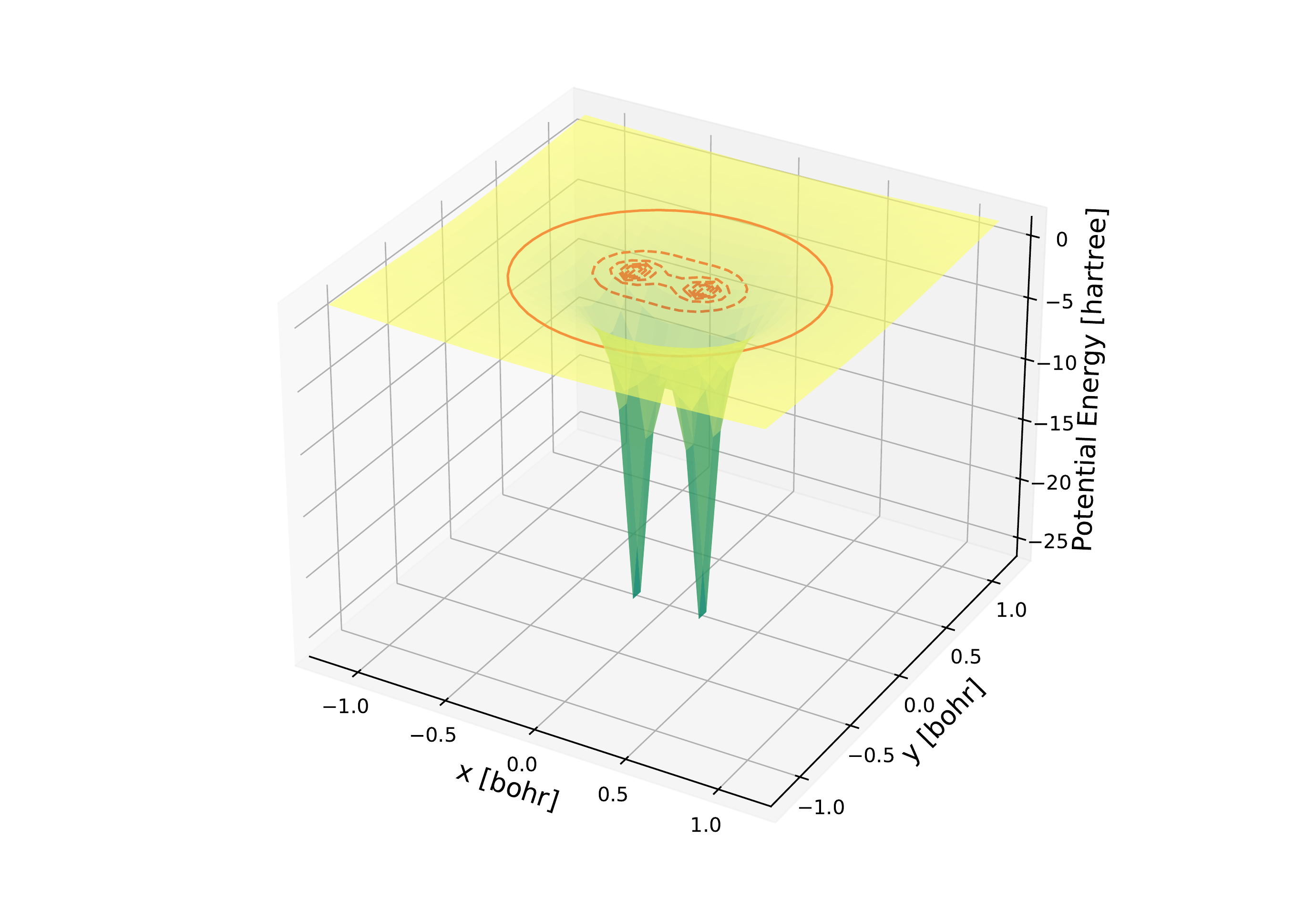}
\caption{Potential energy surface for the H$_2^+$ molecule.}
\label{fig1}
\end{figure}

\begin{figure}
\centering
\includegraphics[width=8.6cm]{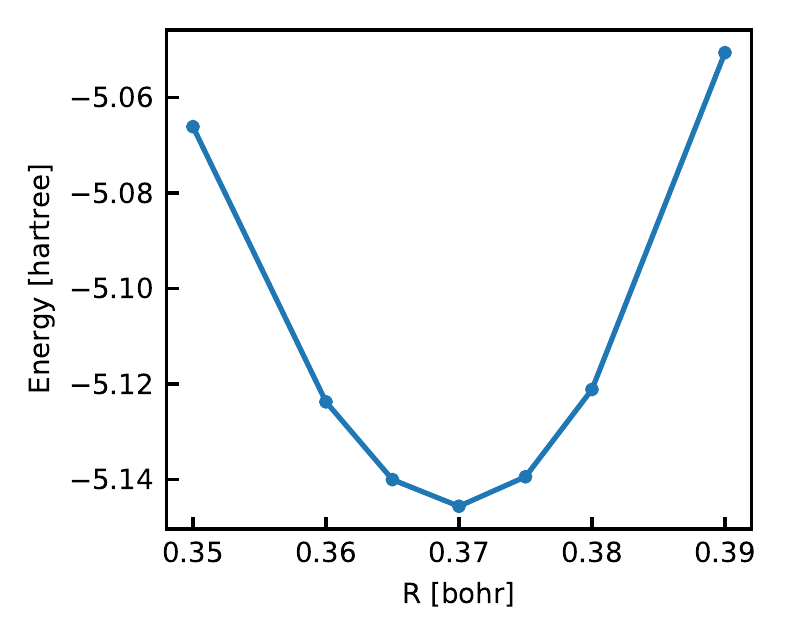}
\caption{Bond lengths and energies for the H$_2^+$ molecule.}
\label{fig2}
\end{figure}

\begin{figure}
\centering
\includegraphics[width=8.6cm]{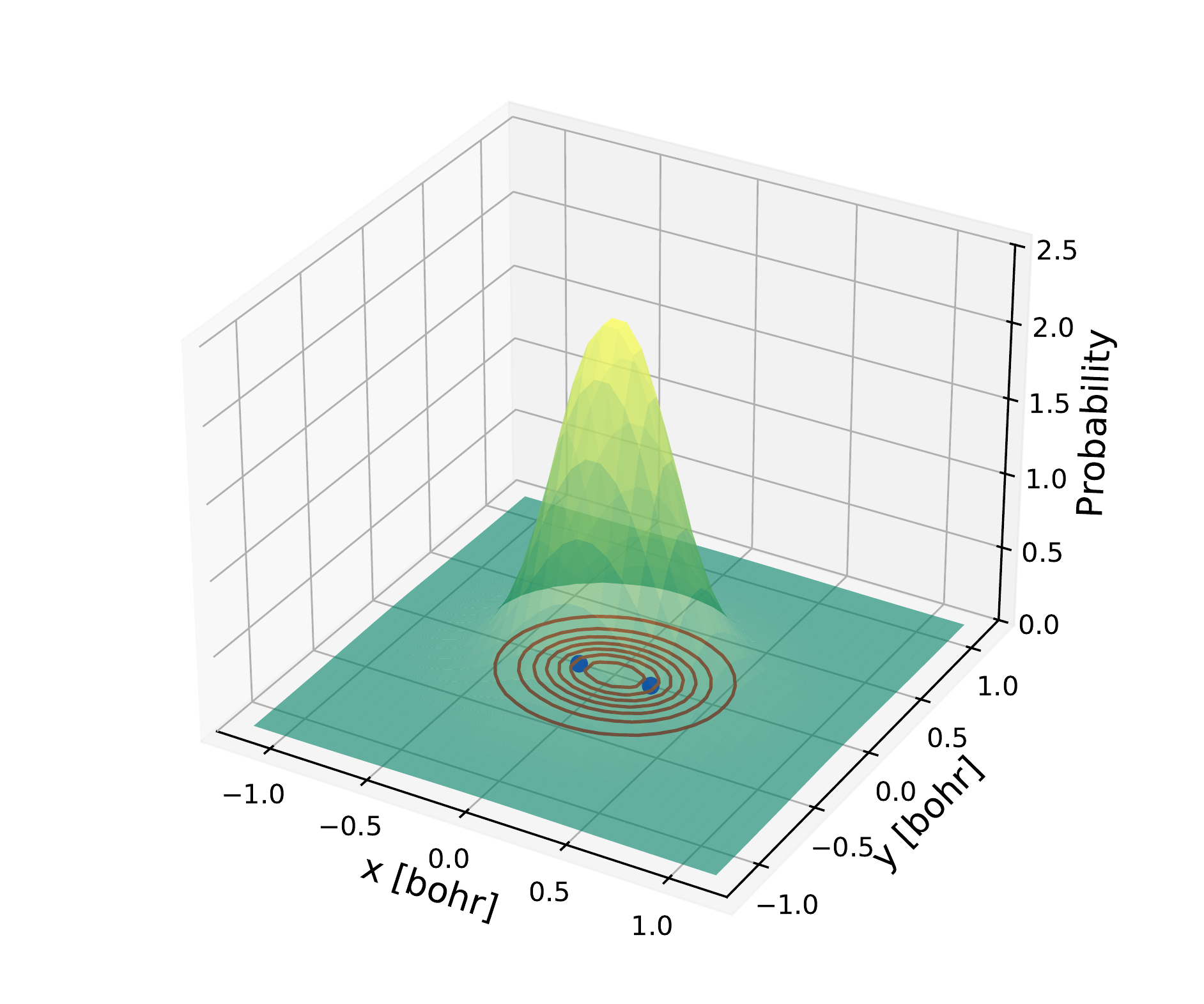}
\caption{Electron probability distribution for the H$_2^+$ molecule.}
\label{fig3}
\end{figure}

For comparison, the 2D H$_2^+ $ molecule was first calculated for the case where the protons were treated as point charges (classical particles).
As a point charge, the electronic state does not change, even if the mass of the proton is changed.
Place two protons at $\vec{R}_1 = (-R/2, 0)$ and $\vec{R}_2 = (R/2, 0) $, and parameterize the internuclear distance $R$.
The electronic state of the ground state was obtained by the imaginary time evolution method. 
The potential energy surface acting on the electron is visualized in FIG. \ref{fig1} in the case of $R = 0.37$ bohr. 
The total energies were calculated with the optimized wave functions while changing $R$ and the potential energy surface was obtained, as shown in FIG. \Ref{fig2}.
From FIG. \Ref{fig2}, the equilibrium internuclear distance when the protons are treated as classical particles is 0.37 bohr. 
The electron probability distribution at this time is shown in FIG. \ref{fig3}. 
Here, the positions of the protons are indicated by blue dots, and the electron probability distribution has a maximum value between the two protons.

\section{Application to larger molecules}
As the size of the molecule increases and the number of degrees of freedom increases, the number of vibrational modes also increases.
As a result, some modes with large fluctuations (e.g., breathing mode) appear, and the fluctuation width of the zero-point vibration also increases, which may affect the number of measurements for the molecular structure optimization by our method.
To estimate how the fluctuation width of the zero-point vibration changes with the molecular size, we consider the one-dimensional model system with $N$ mass points connected by springs. Both ends of the chain of the mass points are assumed to be free. The masses of all the mass points and the spring constants are assumed to be $m$ and $k$, respectively.
The solution of this model can be obtained analytically, and the frequencies of the $n$-th normal modes are given by
\begin{equation}
\omega^{(n)}=2\sqrt{\frac{k}{m}}\sin\frac{n\pi}{N}.
\end{equation}
Here, $n=1,2,\cdots,N-1$.
We are only interested in the low frequency modes with large fluctuation widths, so we assume that $n \ll N$. Then the above equation can be simplified to
\begin{equation}
\omega^{(n)}\simeq 2\sqrt{\frac{k}{m}}\frac{n\pi}{N}.
\end{equation}
The standard deviation of the zero-point fluctuation for the harmonic oscillator with the frequency $\omega$ can be written as $\sigma=\sqrt{\frac{1}{2m\omega}}$,
the standard deviation of the zero-point fluctuation for the lowest frequency mode ($n=1$) can be estimated as 
\begin{equation}
\sigma^{(n=1)}=\sqrt{\frac{1}{2m\omega^{(n=1)}}}\simeq \sqrt{\frac{N}{m\omega_0\pi}},
\label{eq_sig}
\end{equation}
where $\omega_0 = \sqrt{\frac{k}{m}}$ is the frequency of the single string system.
Equation (\ref{eq_sig}) means that the standard deviation of the zero-point fluctuation increases approximately in proportion to $\sqrt{N}$.